\documentclass[prl,superscriptaddress,twocolumn,nofootinbib,showpacs,preprintnumbers,floatfix]{revtex4}

\usepackage{mathrsfs}
\usepackage{amsfonts,amssymb,amsmath}
\usepackage{graphicx,epsfig}
\usepackage{multirow}
\usepackage{verbatim}

\usepackage{slashed}

\declareslashed{}{/}{0}{-0.05}{E}
\declareslashed{}{/}{0}{-0.05}{P}
\declareslashed{}{/}{0}{-0.10}{\Widehat{P}}

\def\fsu5{$\cal{F}$-$SU(5)$}

\def\bfsu5{$\boldsymbol{\mathcal{F}}$-$\boldsymbol{SU(5)}$}

\def\m1half{$M_{1/2}$}
\def\m3half{$M_{3/2}$}
\def\m32{$M_{32}$}

\def\mt2{$M_{T2}$}
\def\x2{$\chi^2$}
\def\2b{$M_{T2}b$}

\def\bs0{$B_S^0 \rightarrow \mu^+ \mu^-$}

\def\bea{\begin{eqnarray}}
\def\eea{\end{eqnarray}}

\def\nnb{\nonumber}

\begin{document}

\title{Inspiration from Intersecting D-branes:\\ General Supersymmetry Breaking Soft Terms in No-Scale ${\cal F}$-$SU(5)$}

\author{Ron De Benedetti}

\affiliation{Department of Chemistry and Physics, Louisiana State University, Shreveport, Louisiana 71115 USA}

\author{Chuang Li}

\affiliation{CAS Key Laboratory of Theoretical Physics, Institute of Theoretical Physics, 
Chinese Academy of Sciences, Beijing 100190, China}

\affiliation{ School of Physical Sciences, University of Chinese Academy of Sciences, 
No.19A Yuquan Road, Beijing 100049, China}

\author{Tianjun Li}

\affiliation{CAS Key Laboratory of Theoretical Physics, Institute of Theoretical Physics, 
Chinese Academy of Sciences, Beijing 100190, China}

\affiliation{ School of Physical Sciences, University of Chinese Academy of Sciences, 
No.19A Yuquan Road, Beijing 100049, China}

\author{Adam Lux}

\affiliation{Department of Physics and Engineering Physics, The University of Tulsa, Tulsa, OK 74104 USA}

\author{James A. Maxin}

\affiliation{Department of Chemistry and Physics, Louisiana State University, Shreveport, Louisiana 71115 USA}

\author{Dimitri V. Nanopoulos}

\affiliation{George P. and Cynthia W. Mitchell Institute for Fundamental Physics and Astronomy, Texas A$\&$M University, College Station, TX 77843, USA}

\affiliation{Astroparticle Physics Group, Houston Advanced Research Center (HARC), Mitchell Campus, Woodlands, TX 77381, USA}

\affiliation{Academy of Athens, Division of Natural Sciences, 28 Panepistimiou Avenue, Athens 10679, Greece}


\begin{abstract}

Motivated by D-brane model building, we evaluate the $\cal{F}$-$SU(5)$ model with additional vector-like particle multiplets, referred to as flippons, within the framework of No-Scale Supergravity with non-vanishing general supersymmetry breaking soft terms at the string scale. The viable phenomenology is uncovered by applying all current experimental constraints, including but not limited to the correct light Higgs boson mass, WMAP and Planck relic density measurements, and several LHC constraints on supersymmetric particle spectra. Four interesting regions of the parameter space arise, as well as mixed scenarios, given by: (i) light stop coannihilation; (ii) pure Higgsino dark matter; (iii) Higgs funnel; and (iv) light stau coannihilation. All regions can generate the observed value of the relic density commensurate with a 125 GeV light Higgs boson mass, with the exception of the relatively small relic density value for the pure Higgsino lightest supersymmetric particle (LSP). This work is concluded by gauging the model against present LHC search constraints and derivation of the final states observable at the LHC for each of these scenarios.

\end{abstract}



\preprint{ACT-02-18, MI-TH-1891}

\maketitle


\section{Introduction}

Successful confirmation of the Standard Model (SM) was celebrated when the lightest CP-even Higgs boson with mass $m_h=125.09\pm 0.24$~GeV was discovered at the LHC in 2012~\cite{Aad:2012tfa, Chatrchyan:2012xdj}. Despite the memorable occasion, severe anomalies persisted in the SM, for instance, the gauge hierarchy problem, conspicuous absence of gauge coupling unification, and lack of a plausible dark matter candidate, just to highlight a few. Beyond the SM (BSM) theories, and supersymmetry in particular, can rescue high-energy physics from these SM deficiencies. Supersymmetry (SUSY) can solve the gauge hierarchy problem and produce gauge coupling unification. In supersymmetric SMs (SSMs), the large top quark Yukawa coupling can radiatively break electroweak (EW) gauge symmetry. The lightest supersymmetric particle (LSP) neutralino ($\widetilde{\chi}_1^0$) can serve as a viable dark matter candidate in SSMs with $R$-parity. Of particular significance, gauge coupling unification strongly suggests Grand Unified Theories (GUTs), and SUSY GUTs can be elegantly constructed from superstring theory. Supersymmetry thus provides rather auspicious new physics beyond the SM, beautifully merging low energy phenomenology with high-energy fundamental physics. Despite the successful intervention SUSY can inject into high-energy physics, it remains considerably challenging to obtain a light Higgs boson mass around 125 GeV in the Minimal SSM (MSSM) without embracing either multi-TeV top squarks with small mixing or TeV-scale top squarks with large mixing~\cite{Carena:2011aa}. Compounding the effort are the strong constraints on the SSM viable parameter space established by the LHC SUSY searches. Principally among those LHC constraints are the masses for the gluino (${\tilde g}$) and light stop (${\tilde t}_1$), where the nominal exclusion curves imply the masses are heavier than about 1.9 TeV and 900~GeV, respectively~\cite{WAdam-ICHEP}, indicating that an electroweak fine-tuning problem may potentially lurk in SUSY.

The aforementioned threats to developing effective BSM constructions notwithstanding, string theory perseveres as one of the most promising theories for quantum gravity. As opposed to the conventional unification at the GUT scale realized in the typical SUSY GUT, we proposed the testable flipped $SU(5)\times U(1)_X$ models~\cite{Barr:1981qv,Derendinger:1983aj,Antoniadis:1987dx} with additional TeV-scale vector-like particles~\cite{Jiang:2006hf}, which we colorfully dubbed flippons, to obtain gauge coupling unification at the $string~scale$. Subsequently, we further constructed this class of flipped $SU(5)$ models from local F-theory model building~\cite{Jiang:2008yf, Jiang:2009za}. These flipped $SU(5)$ models with extra vector-like multiplets can be realized in free-fermionic string constructions also~\cite{LNY}, hence we denoted them ${\cal F}$-$SU(5)$. Let's enumerate the ``Miracles''~\cite{Li:2011ab} of the flippons in ${\cal F}$-$SU(5)$:

\noindent  (1) The lightest CP-even Higgs boson mass can be easily lifted to 125~GeV due to one-loop contributions from the Yukawa couplings between the flippons and Higgs fields~\cite{Huo:2011zt, Li:2011ab}.

\noindent (2)  It is well-known that dimension-five proton decays mediated by colored Higgsinos are highly suppressed due to the missing partner mechanism and TeV-scale $\mu$ term in the flipped $SU(5)\times U(1)_X$ models. In ${\cal F}$-$SU(5)$, the $SU(3)_C\times SU(2)_L$ gauge couplings do in fact unify at the traditional GUT scale while the two unified gauge couplings increase as a result of the vector-like particle contributions~\cite{Li:2009fq, Li:2010dp}. Therefore, dimension-six proton decays via heavy gauge boson exchanges are within the reach of the future proton decay experiments such as the Hyper-Kamiokande experiment. More concisely, ${\cal F}$-$SU(5)$ models differ from the minimal flipped $SU(5)\times U(1)_X$ model since the proton lifetime in the minimal model is too lengthy for future proton decay experiments.

\noindent (3) The lightest neutralino serves as the LSP and is lighter than the light stau attributable to the longer running of the Renormalization Group Equations (RGEs) in No-Scale supergravity~\cite{Cremmer:1983bf}, allowing the LSP neutralino to prevail as a dark matter candidate~\cite{Li:2010ws, Li:2010mi, Li:2011xua}. More acutely, No-Scale ${\cal F}$-$SU(5)$ yields the uncommon mass hierarchy $M({\tilde t}_1) < M({\tilde g}) < M({\tilde q}) $ of a light stop and gluino both substantially lighter than all other squarks (${\tilde q}$)~\cite{Li:2010ws, Li:2010mi, Li:2011xua}. The net effect of this rare SUSY spectrum is the production of four top quarks, leading to large multijet events at the LHC~\cite{Maxin:2011hy}.

\noindent (4) An alliance between No-Scale supergravity and the Giudice-Masiero (GM) mechanism~\cite{Giudice:1988yz} allows the SUSY electroweak fine-tuning problem to be resolved naturally~\cite{Leggett:2014mza, Leggett:2014hha}. 

Several prior No-Scale \fsu5 analyses have intimately examined vanishing SUSY breaking soft terms with the exception of a single unified gaugino parameter $M_{1/2}$ (see, for example, Refs.~\cite{Li:2011ab,Li:2016bww,Li:2017kcq}). To complement those prior sweeping studies, we now consider in this paper non-zero general SUSY breaking soft terms in No-Scale ${\cal F}$-$SU(5)$, partially inspired by D-brane model building~\cite{Chen:2006ip}. The low-energy particle spectra will be examined that are consistent with all the current experimental constraints, and interesting diverse regions of the parameter space that can generate the observed dark matter relic density and correct light Higgs boson mass, in addition to several other currently operating experiments, will be identified and analyzed discretely to derive low-energy phenomenology. Finally, benchmarks models will be classified and branching fractions computed to itemize the final states observable at the LHC.

\section{The \bfsu5 Model}

Here we only briefly review the minimal flipped $SU(5)$ model~\cite{Barr:1981qv,Derendinger:1983aj,Antoniadis:1987dx}, where the gauge group $SU(5)\times U(1)_{X}$ can be embedded into the $SO(10)$ model. More comprehensive discussions of the minimal flipped $SU(5)$ model can be found in Refs.~\cite{Maxin:2011hy,Li:2011ab,Li:2013naa,Leggett:2014hha,Li:2016bww} and references therein. We define the generator $U(1)_{Y'}$ in $SU(5)$ as 
\bea 
T_{\rm U(1)_{Y'}}={\rm diag} \left(-\frac{1}{3}, -\frac{1}{3}, -\frac{1}{3},
 \frac{1}{2},  \frac{1}{2} \right)~,~\,
\label{u1yp}
\eea
and subsequently the hypercharge is given by
\bea
Q_{Y} = \frac{1}{5} \left( Q_{X}-Q_{Y'} \right).
\label{ycharge}
\eea
We have three families of the SM fermions whose quantum numbers under $SU(5)\times U(1)_{X}$ are respectively
\bea
F_i={\mathbf{(10, 1)}},~ {\bar f}_i={\mathbf{(\bar 5, -3)}},~
{\bar l}_i={\mathbf{(1, 5)}},
\label{smfermions}
\eea
where $i=1, 2, 3$. The SM particle assignments in $F_i$, ${\bar f}_i$ and ${\bar l}_i$ are
\bea
F_i=(Q_i, D^c_i, N^c_i),~{\overline f}_i=(U^c_i, L_i),~{\overline l}_i=E^c_i~,~
\label{smparticles}
\eea
where $Q_i$, $U^c_i$, $D^c_i$,  $L_i$, $E^c_i$ and $N^c_i$ are the left-handed quark doublets, right-handed up-type quarks, down-type quarks, left-handed lepton doublets, right-handed charged leptons, and neutrinos, respectively. Generation of the heavy right-handed neutrino masses is accomplished by introduction of three SM singlets $\phi_i$.

Breaking the GUT and electroweak gauge symmetries is achieved via introduction of the two pairs of Higgs representations
\begin{eqnarray}
H&=&{\mathbf{(10, 1)}},~{\overline{H}}={\mathbf{({\overline{10}}, -1)}}, \nonumber \\
h&=&{\mathbf{(5, -2)}},~{\overline h}={\mathbf{({\bar {5}}, 2)}}.
\label{Higgse1}
\end{eqnarray}
The states in the $H$ multiplet are labeled by the same symbols as in the $F$ multiplet, and for ${\overline H}$ we merely add ``bar'' above the fields. Explicitly, the Higgs particles are
\bea
H=(Q_H, D_H^c, N_H^c)~,~
{\overline{H}}= ({\overline{Q}}_{\overline{H}}, {\overline{D}}^c_{\overline{H}}, 
{\overline {N}}^c_{\overline H})~,~\,
\label{Higgse2}
\eea
\bea
h=(D_h, D_h, D_h, H_d)~,~
{\overline h}=({\overline {D}}_{\overline h}, {\overline {D}}_{\overline h},
{\overline {D}}_{\overline h}, H_u)~,~\,
\label{Higgse3}
\eea
where $H_d$ and $H_u$ are one pair of Higgs doublets in the MSSM.

The $SU(5)\times U(1)_{X}$ gauge symmetry is broken down to the SM gauge symmetry by the following Higgs superpotential at the GUT scale
\bea
{\it W}_{\rm GUT}=\lambda_1 H H h + \lambda_2 {\overline H} {\overline H} {\overline
h} + \Phi ({\overline H} H-M_{\rm H}^2)~.~ 
\label{spgut} 
\eea
Only one F-flat and D-flat direction exists, which can be rotated along the $N^c_H$ and ${\overline {N}}^c_{\overline H}$ directions. Hence, 
we obtain $<N^c_H>=<{\overline {N}}^c_{\overline H}>=M_{\rm H}$. Furthermore, the superfields $H$ and ${\overline H}$ are ``eaten'' and acquire substantial masses via the supersymmetric Higgs mechanism, with the exception of $D_H^c$ and ${\overline {D}}^c_{\overline H}$. Additionally, the superpotential terms $ \lambda_1 H H h$ and $ \lambda_2 {\overline H} {\overline H} {\overline h}$ couple $D_H^c$ and ${\overline {D}}^c_{\overline H}$ respectively with $D_h$ and ${\overline {D}}_{\overline h}$, forming massive eigenstates with masses $2 \lambda_1 <N_H^c>$ and $2 \lambda_2 <{\overline {N}}^c_{\overline H}>$. As a result, we naturally experience the doublet-triplet splitting due to the missing partner mechanism~\cite{Antoniadis:1987dx}. The triplets in $h$ and ${\overline h}$ though only have a small mixing through the $\mu$ term, and as such, the colored Higgsino-exchange mediated proton decay is negligible, {\it i.e.}, there is no dimension-5 proton decay problem. 

To realize string-scale gauge coupling unification~\cite{Jiang:2006hf, Jiang:2008yf, Jiang:2009za}, we introduce the following vector-like particles (flippons) at the TeV scale
\begin{eqnarray}
&& XF ={\mathbf{(10, 1)}}~,~{\overline{XF}}={\mathbf{({\overline{10}}, -1)}}~,~\nnb \\
&& Xl={\mathbf{(1, -5)}}~,~{\overline{Xl}}={\mathbf{(1, 5)}}~.~\,
\end{eqnarray}
The particle content from the decompositions of $XF$, ${\overline{XF}}$, $Xl$, and ${\overline{Xl}}$ under the SM gauge symmetry are
\begin{eqnarray}
&& XF = (XQ, XD^c, XN^c)~,~ {\overline{XF}}=(XQ^c, XD, XN)~,~\nnb \\
&& Xl= XE~,~ {\overline{Xl}}= XE^c~.~
\end{eqnarray}
The quantum numbers for the extra vector-like particles under the $SU(3)_C \times SU(2)_L \times U(1)_Y$ gauge symmetry are
\begin{eqnarray}
&& XQ={\mathbf{(3, 2, \frac{1}{6})}}~,~
XQ^c={\mathbf{({\bar 3}, 2,-\frac{1}{6})}} ~,~\\
&& XD={\mathbf{({3},1, -\frac{1}{3})}}~,~
XD^c={\mathbf{({\bar 3},  1, \frac{1}{3})}}~,~\\
&& XN={\mathbf{({1},  1, {0})}}~,~
XN^c={\mathbf{({1},  1, {0})}} ~,~\\
&& XE={\mathbf{({1},  1, {-1})}}~,~
XE^c={\mathbf{({1},  1, {1})}}~.~\,
\label{qnum}
\end{eqnarray}
The superpotential is
\bea 
{ W}_{\rm Yukawa} &=&  y_{ij}^{D}
F_i F_j h + y_{ij}^{U \nu} F_i  {\overline f}_j {\overline
h}+ y_{ij}^{E} {\overline l}_i  {\overline f}_j h  
\nnb \\ &&
+ \mu h {\overline h}
+ y_{ij}^{N} \phi_i {\overline H} F_j +M_{ij}^{\phi} \phi_i \phi_j
\nnb \\ &&
+ y_{XF} XF XF h + y_{\overline{XF}} {\overline{XF}} {\overline{XF}} {\overline h}
\nnb \\ &&
+ M_{XF} {\overline{XF}}  XF + M_{Xl} {\overline{Xl}}  Xl
~,~\,
\label{potgut}
\eea
and after the $SU(5)\times U(1)_X$ gauge symmetry is broken down to the SM gauge symmetry, the above superpotential gives 
\bea 
{ W_{SSM}}&=&
y_{ij}^{D} D^c_i Q_j H_d+ y_{ji}^{U \nu} U^c_i Q_j H_u
+ y_{ij}^{E} E^c_i L_j H_d \nnb \\ &&
+  y_{ij}^{U \nu} N^c_i L_j H_u  +  \mu H_d H_u+ y_{ij}^{N} 
\langle {\overline {N}}^c_{\overline H} \rangle \phi_i N^c_j
\nnb \\ &&
+ y_{XF} XQ XD^c H_d + y_{\overline{XF}} XQ^c XD H_u
\nnb \\ &&
+M_{XF}\left(XQ^c XQ + XD^c XD\right) 
\nnb \\ &&
+ M_{Xl} XE^c  XE+M_{ij}^{\phi} \phi_i \phi_j
\nnb \\ &&
 + \cdots (\textrm{decoupled below $M_{GUT}$}). 
\label{poten1}
\eea
where $y_{ij}^{D}$, $y_{ij}^{U \nu}$, $y_{ij}^{E}$, $y_{ij}^{N}$, $y_{XF} $, and $y_{\overline{XF}}$ are Yukawa couplings, $\mu$ is the bilinear Higgs mass term, and $M_{ij}^{\phi}$, $M_{XF} $ and $M_{Xl}$ are masses for new particles. These new particles are of course our flippons, however, we shall not explicitly compute the masses $M_{ij}^{\phi}$, $M_{XF}$ ,and $M_{Xl}$ here in this work, reserving that project for a future date.  Nonetheless, we do implement a common mass decoupling scale $M_V$ for the flippon vector-like particles. Current LHC constraints on vector-like $T$ and $B$ quarks~\cite{atlas-vectorlike} provide lower limits of around 855~GeV for our $(XQ, ~XQ^c)$ vector-like flippons and 735~GeV for our $(XD, ~XD^c)$ vector-like flippons. Accordingly, we establish our lower $M_V$ limit at $M_V \ge 855$~GeV to ensure complete coverage of all experimentally viable flippon masses in our analysis.

\begin{table*}[htp]
  \centering
  \scriptsize
  \caption{The \fsu5 general SUSY breaking soft terms, in addition to the vector-like flippon decoupling scale $M_V$, the low energy ratio of Higgs vacuum expectation values (VEVs) tan$\beta$, and top quark mass $M_t$, for the four regions of the model space we study in this work. Each benchmark point is identified with an alphabetical label in order to link the data in TABLE~\ref{tab:spectra1} with the data in TABLES~\ref{tab:spectra2} - \ref{tab:spectra3}. All masses are in GeV.}
\label{tab:spectra1}
\begin{tabular}{|c|c||c|c|c|c|c|c|c|c|c|c|c|c|} \hline
${\rm Model}$ & $ {\rm benchmark} $ & $M_5$ & $M_{1{\rm X}}$  &  $M_{U^c L}$  &  $M_{E^c}$  &  $M_{Q D^c N^c}$  &  $M_H$  &  $A_{\tau}$  &  $A_t$  &  $A_b$  &   $M_V$  &  $ {\rm tan}\beta $  &  $M_t$ \\ \hline \hline
$	{\rm Stop~Coannihilation}	$ & $	{\rm A}	$ & $	1650	$ & $	3125	$ & $	2142	$ & $	1158	$ & $	3125	$ & $	175	$ & $	3125	$ & $	-4850	$ & $	1250	$ & $	16250	$ & $	17.58	$ & $	174.43	$	\\ \hline
$	{\rm Stop~Coannihilation}	$ & $	{\rm B}	$ & $	1800	$ & $	3275	$ & $	325	$ & $	3275	$ & $	3275	$ & $	1308	$ & $	325	$ & $	-4550	$ & $	-1500	$ & $	16550	$ & $	42.74	$ & $	171.85	$	\\ \hline
$	{\rm Stop~Coannihilation}	$ & $	{\rm C}	$ & $	2325	$ & $	3800	$ & $	850	$ & $	2817	$ & $	2325	$ & $	3800	$ & $	2600	$ & $	-3500	$ & $	2600	$ & $	17600	$ & $	24.33	$ & $	174.56	$	\\ \hline \hline
$	{\rm Pure~Higgsino}	$ & $	{\rm D}	$ & $	1700	$ & $	4650	$ & $	3667	$ & $	3667	$ & $	1700	$ & $	2683	$ & $	1250	$ & $	4300	$ & $	-1800	$ & $	4055	$ & $	32.83	$ & $	173.43	$	\\ \hline
$	{\rm Pure~Higgsino}	$ & $	{\rm E}	$ & $	2050	$ & $	5000	$ & $	3525	$ & $	5000	$ & $	5000	$ & $	5000	$ & $	-1100	$ & $	1950	$ & $	-1100	$ & $	4755	$ & $	48.16	$ & $	174.80	$	\\ \hline
$	{\rm Pure~Higgsino}	$ & $	{\rm F}	$ & $	2375	$ & $	3850	$ & $	2867	$ & $	3850	$ & $	2375	$ & $	3850	$ & $	-350	$ & $	2700	$ & $	-350	$ & $	2455	$ & $	36.66	$ & $	173.26	$	\\ \hline \hline
$	{\rm Higgs~Funnel}	$ & $	{\rm G}	$ & $	1500	$ & $	2975	$ & $	4450	$ & $	2483	$ & $	2975	$ & $	2483	$ & $	550	$ & $	3600	$ & $	-2500	$ & $	3655	$ & $	30.83	$ & $	173.38	$	\\ \hline
$	{\rm Higgs~Funnel}	$ & $	{\rm H}	$ & $	1750	$ & $	3225	$ & $	3225	$ & $	2242	$ & $	275	$ & $	1258	$ & $	275	$ & $	1450	$ & $	-1600	$ & $	16450	$ & $	30.41	$ & $	173.14	$	\\ \hline
$	{\rm Higgs~Funnel}	$ & $	{\rm I}	$ & $	2300	$ & $	3775	$ & $	3775	$ & $	2792	$ & $	825	$ & $	2792	$ & $	3775	$ & $	2550	$ & $	-500	$ & $	9928	$ & $	35.91	$ & $	173.25	$	\\ \hline \hline
$	{\rm Stau~Coannihilation}	$ & $	{\rm J}	$ & $	1600	$ & $	1600	$ & $	125	$ & $	1108	$ & $	1600	$ & $	1108	$ & $	125	$ & $	-1900	$ & $	1150	$ & $	16150	$ & $	28.91	$ & $	173.11	$	\\ \hline
$	{\rm Stau~Coannihilation}	$ & $	{\rm K}	$ & $	2175	$ & $	3650	$ & $	700	$ & $	1683	$ & $	700	$ & $	2667	$ & $	-750	$ & $	2300	$ & $	2300	$ & $	9678	$ & $	46.49	$ & $	171.92	$	\\ \hline
$	{\rm Stau~Coannihilation}	$ & $	{\rm L}	$ & $	2625	$ & $	4100	$ & $	1150	$ & $	3117	$ & $	2625	$ & $	1150	$ & $	-2900	$ & $	-2900	$ & $	3200	$ & $	2955	$ & $	50.99	$ & $	172.01	$	\\ \hline \end{tabular}
\end{table*}

\begin{table*}[htp]
  \centering
  \scriptsize
  \caption{Relevant SUSY spectrum masses for the \fsu5 general SUSY breaking soft terms of TABLE~\ref{tab:spectra1}. The soft SUSY breaking terms that generate each of these spectra can be identified by the alphabetical label. A $\dagger$ symbol in the light Higgs boson mass $m_h$ column represents the theoretically computed value consisting of $only$ the 1-loop and 2-loop SUSY contributions, primarily from the coupling to the light stop, but does $not$ include any vector-like flippon contributions, whereas a $\dagger \dagger$ symbol represents the 1-loop and 2-loop SUSY contributions $plus$ the maximum vector-like flippon contribution. All masses are in GeV.}
\label{tab:spectra2}
\begin{tabular}{|c|c||c|c|c|c|c|c|c|c|c|} \hline
${\rm Model}$ & $ {\rm Benchmark} $ & $M({\tilde{\chi}^0_1})$ & $M({\tilde{\chi}^0_2})$  &  $M({\tilde{\chi}^\pm_1})$  &  $M({\tilde{\tau}_1^\pm})$  &  $M({\tilde{t}_1})$  &  $M({\tilde{u}_R})$  &  $M({\tilde{g}})$  &  $m_h$  &  $M({H^0/A^0})$ \\ \hline \hline
$	{\rm Stop~Coannihilation}	$ & $	{\rm A}	$ & $	693	$ & $	781	$ & $	781	$ & $	1080	$ & $	729	$ & $	3759	$ & $	2225	$ & $	127.34^{\dagger}	$ & $	4484	$	\\ \hline
$	{\rm Stop~Coannihilation}	$ & $	{\rm B}	$ & $	728	$ & $	850	$ & $	850	$ & $	2058	$ & $	766	$ & $	3727	$ & $	2397	$ & $	125.85^{\dagger}	$ & $	3927	$	\\ \hline
$	{\rm Stop~Coannihilation}	$ & $	{\rm C}	$ & $	862	$ & $	1105	$ & $	1105	$ & $	2418	$ & $	895	$ & $	4370	$ & $	3020	$ & $	127.60^{\dagger}	$ & $	5134	$	\\ \hline \hline
$	{\rm Pure~Higgsino}	$ & $	{\rm D}	$ & $	188	$ & $	-198	$ & $	192	$ & $	3729	$ & $	3012	$ & $	4695	$ & $	2246	$ & $	126.73^{\dagger \dagger}	$ & $	1268	$	\\ \hline
$	{\rm Pure~Higgsino}	$ & $	{\rm E}	$ & $	411	$ & $	-421	$ & $	415	$ & $	3241	$ & $	3586	$ & $	5591	$ & $	2726	$ & $	126.21^{\dagger}	$ & $	1794	$	\\ \hline
$	{\rm Pure~Higgsino}	$ & $	{\rm F}	$ & $	374	$ & $	-383	$ & $	378	$ & $	3482	$ & $	3699	$ & $	5293	$ & $	3068	$ & $	125.99^{\dagger}	$ & $	2811	$	\\ \hline \hline
$	{\rm Higgs~Funnel}	$ & $	{\rm G}	$ & $	621	$ & $	679	$ & $	679	$ & $	3326	$ & $	3556	$ & $	4909	$ & $	2028	$ & $	127.18^{\dagger \dagger}	$ & $	1316	$	\\ \hline
$	{\rm Higgs~Funnel}	$ & $	{\rm H}	$ & $	717	$ & $	824	$ & $	824	$ & $	3045	$ & $	2729	$ & $	4172	$ & $	2337	$ & $	124.87^{\dagger \dagger}	$ & $	1449	$	\\ \hline
$	{\rm Higgs~Funnel}	$ & $	{\rm I}	$ & $	840	$ & $	1068	$ & $	1068	$ & $	3435	$ & $	3564	$ & $	5239	$ & $	2989	$ & $	124.67^{\dagger}	$ & $	1739	$	\\ \hline \hline
$	{\rm Stau~Coannihilation}	$ & $	{\rm J}	$ & $	362	$ & $	749	$ & $	749	$ & $	365	$ & $	1248	$ & $	2937	$ & $	2121	$ & $	126.77^{\dagger}	$ & $	3110	$	\\ \hline
$	{\rm Stau~Coannihilation}	$ & $	{\rm K}	$ & $	799	$ & $	918	$ & $	912	$ & $	802	$ & $	2729	$ & $	3985	$ & $	2805	$ & $	125.62^{\dagger \dagger}	$ & $	1781	$	\\ \hline
$	{\rm Stau~Coannihilation}	$ & $	{\rm L}	$ & $	870	$ & $	1177	$ & $	1177	$ & $	876	$ & $	2632	$ & $	5148	$ & $	3278	$ & $	126.43^{\dagger}	$ & $	4179	$	\\ \hline
\end{tabular}
\end{table*}

\begin{table*}[htp]
  \centering
  \scriptsize
  \caption{Relic density ($\Omega h^2$), rare decay processes ($\Delta a_\mu$, $Br(b \to s\gamma)$, $Br(B_s^0 \to \mu^+ \mu^-)$, rescaled dark matter direct-detection cross sections ($ \sigma_{SI}$, $\sigma_{SD}$), and $p \to e^+ \pi^0$ proton decay rates ($\tau_p$) for the \fsu5 general SUSY breaking soft terms of TABLE~\ref{tab:spectra1}. The soft SUSY breaking terms that generate each of these values can be identified by the alphabetical label. The $\sigma_{SI}$ and $\sigma_{SD}$ cross-sections have been rescaled in accordance with Eq.~(\ref{eq:omega}). The numerical values given for $\Delta a_{\mu}$ are $\times 10^{-10}$, $Br(b \rightarrow s \gamma)$ are $\times 10^{-4}$,  $Br(B_s^0 \rightarrow \mu^+ \mu^-)$ are $\times 10^{-9}$, rescaled spin-independent cross-sections $\sigma_{SI}$ are $\times 10^{-11}$~pb, rescaled spin-dependent cross-sections $\sigma_{SD}$ are $\times 10^{-9}$~pb, and proton decay rate $\tau_p$ are $\times 10^{35}$~yrs.}
\label{tab:spectra3}
\begin{tabular}{|c|c||c|c|c|c|c|c|c|c|} \hline
${\rm Model}$ & $ {\rm Benchmark} $ & $\Omega h^2$  &  $\Delta a_\mu$  &  $Br(b \to s\gamma)$  &  $Br(B_s^0 \to \mu^+ \mu^-)$  &  $ \sigma_{SI}$  &  $\sigma_{SD}$  & $\tau_p$ & $ {\rm LSP}$ \\ \hline \hline
$	{\rm Stop~Coannihilation}	$ & $	{\rm A}	$ & $	0.1115	$ & $	0.40	$ & $	3.56	$ & $	3.16	$ & $	0.1	$ & $	0.1	$  & $	3.9	$ & $	>99\%~{\rm Bino}	$	\\ \hline
$	{\rm Stop~Coannihilation}	$ & $	{\rm B}	$ & $	0.1219	$ & $	1.02	$ & $	3.45	$ & $	3.57	$ & $	0.1	$ & $	0.1	$  & $	5.9	$ & $	>99\%~{\rm Bino}	$	\\ \hline
$	{\rm Stop~Coannihilation}	$ & $	{\rm C}	$ & $	0.1045	$ & $	0.60	$ & $	3.47	$ & $	3.20	$ & $	0.6	$ & $	1.0	$  & $	2.7	$ & $	>99\%~{\rm Bino}	$	\\ \hline \hline
$	{\rm Pure~Higgsino}	$ & $	{\rm D}	$ & $	0.0052	$ & $	0.74	$ & $	3.72	$ & $	2.90	$ & $	34	$ & $	1471	$  & $	0.3	$ & $	>99\%~{\rm Higgsino}	$	\\ \hline
$	{\rm Pure~Higgsino}	$ & $	{\rm E}	$ & $	0.0198	$ & $	0.99	$ & $	3.64	$ & $	3.08	$ & $	125	$ & $	1484	$  & $	0.8	$ & $	>98\%~{\rm Higgsino}	$	\\ \hline
$	{\rm Pure~Higgsino}	$ & $	{\rm F}	$ & $	0.0170	$ & $	0.93	$ & $	3.57	$ & $	3.04	$ & $	77	$ & $	1305	$  & $	0.4	$ & $	>98\%~{\rm Higgsino}	$	\\ \hline \hline
$	{\rm Higgs~Funnel}	$ & $	{\rm G}	$ & $	0.1194	$ & $	0.57	$ & $	3.71	$ & $	3.10	$ & $	15	$ & $	40.8	$  & $	2.0	$ & $	>99\%~{\rm Bino}	$	\\ \hline
$	{\rm Higgs~Funnel}	$ & $	{\rm H}	$ & $	0.1193	$ & $	0.76	$ & $	3.61	$ & $	3.38	$ & $	6.4	$ & $	13.2	$  & $	2.6	$ & $	>99\%~{\rm Bino}	$	\\ \hline
$	{\rm Higgs~Funnel}	$ & $	{\rm I}	$ & $	0.1210	$ & $	0.77	$ & $	3.60	$ & $	3.27	$ & $	46	$ & $	112	$  & $	1.8	$ & $	>99\%~{\rm Bino}	$	\\ \hline \hline
$	{\rm Stau~Coannihilation}	$ & $	{\rm J}	$ & $	0.1189	$ & $	1.81	$ & $	3.44	$ & $	3.26	$ & $	0.3	$ & $	1.1	$  & $	2.9	$ & $	>99\%~{\rm Bino}	$	\\ \hline
$	{\rm Stau~Coannihilation}	$ & $	{\rm K}	$ & $	0.1202	$ & $	2.51	$ & $	3.47	$ & $	3.54	$ & $	940	$ & $	2746	$  & $	1.3	$ & $	>93\%~{\rm Bino}	$	\\ \hline
$	{\rm Stau~Coannihilation}	$ & $	{\rm L}	$ & $	0.1176	$ & $	0.79	$ & $	3.49	$ & $	3.65	$ & $	0.1	$ & $	0.1	$  & $	4.4	$ & $	>99\%~{\rm Bino}	$	\\ \hline
\end{tabular}
\end{table*}

As summarized in the prior section, the split-unification of flipped $SU(5)$~\cite{Barr:1981qv,Derendinger:1983aj,Antoniadis:1987dx} provides for fundamental GUT scale Higgs representations (not adjoints), natural doublet-triplet splitting, suppression of dimension-five proton decay~\cite{Harnik:2004yp}, and a two-step see-saw mechanism for neutrino masses~\cite{Ellis:1992nq,Ellis:1993ks}. Additions to the one-loop gauge $\beta$-function coefficients $b_i$ to include contributions from the vector-like flippon multiplets induce the necessary flattening of the $SU(3)$ Renormalization Group Equation (RGE) running ($b_3 = 0$)~\cite{Li:2010ws}, translating into a clear separation between the primary $SU(3)_C \times SU(2)_L$ unification near $10^{16}$~GeV and the secondary $SU(5) \times U(1)_X$ unification at around $5 \times 10^{17}$~GeV, which we define as the $M_{\cal F}$ scale, thus elevating unification to near the Planck mass. At the primary $SU(3)_C \times SU(2)_L$ unification near $10^{16}$~GeV, the $M2$ and $M3$ gaugino mass terms are unified into a single term that we refer to as $M5$~\cite{Li:2010rz}, where $M5 = M2 = M3$ between the primary unification around $10^{16}$~GeV and the secondary $SU(5) \times U(1)_X$ unification near $5 \times 10^{17}$~GeV~\cite{Li:2010ws}. The $M1$ gaugino mass term runs up to the secondary $SU(5) \times U(1)_X$ unification at $M_{\cal F}$ and unifies with $M5$, by means of a slight shift due to $U(1)_X$ flux effects~\cite{Li:2010rz} between the primary unification around $10^{16}$~GeV and the secondary $SU(5) \times U(1)_X$ unification at $M_{\cal F}$~\cite{Li:2010ws}. Given this shift, we refer to the $M1$ gaugino mass term above the primary unification around $10^{16}$~GeV as $M_{1X}$~\cite{Li:2010rz}.  The resulting baseline extension for logarithmic running of the No-Scale boundary conditions permits sufficient scale for natural dynamic evolution down to viable phenomenology at the electroweak scale. It is this associated flattening of the color-charged gaugino mass scale that produces the distinctive mass texture of $M(\widetilde{t}_1) < M(\widetilde{g}) < M(\widetilde{q})$, generating a light stop and gluino that are lighter than all other squarks~\cite{Li:2011ab}. 

Therefore, the most general supersymmetry breaking soft terms at the scale $M_{\cal F}$ in our ${\cal F}$-$SU(5)$ model are $M_5$, $M_{1X}$, $M_{U^c L}$, $M_{E^c}$, $M_{Q D^c N^c}$, $M_{H_u}$,  $M_{H_d}$, $A_{\tau}$, $A_t$, and $A_b$. General supersymmetry breaking soft terms of these type are partially inspired by D-brane model building~\cite{Chen:2006ip}, where $F_i$, ${\overline f}_i$, ${\overline l}_i$, and $h/{\overline h}$ arise from intersections of different stacks of D-branes. Consequently, the corresponding supersymmetry breaking soft mass terms and trilinear $A$ terms will be different, while $M_{H_u}$ is equal to $M_{H_d}$. Even though the Yukawa terms $H H h$ and ${\overline H} {\overline H} {\overline h}$ of Eq. (\ref{spgut}) and $F_i F_j h$, $XF XF h$, and $\overline{XF} \overline{XF} {\overline h}$ of Eq.~(\ref{potgut}) are forbidden by the anomalous global $U(1)$ symmetry of $U(5)$, we might generate these Yukawa terms from high-dimensional operators or instanton effects. Unlike the $SU(5)$ models, the Yukawa term $F_i F_j h$ in the ${\cal F}$-$SU(5)$ model gives down-type quark masses, so their Yukawa couplings can be small and can be generated via high-dimensional operators or instanton effects.

\section{Numerical Methodology}

The general SUSY breaking soft terms $M_5$, $M_{1X}$, $M_{U^c L}$, $M_{E^c}$, $M_{Q D^c N^c}$,  $M_{H_u} = M_{H_d} = M_H$, $A_{\tau}$, $A_t$, and $A_b$ are applied at the $M_{\cal F}$ scale near $M_{\cal F} \simeq 5 \times 10^{17}$~GeV (in contrast to the traditional GUT scale of about $10^{16}$~GeV in the MSSM), in addition to tan$\beta$, the vector-like flippon mass decoupling scale $M_V$, and the top quark mass $M_t$. The parameter space is sampled within the limits $100 \le M_5 \le 5000$~GeV, $100 \le M_{1X} \le 5000$~GeV, $100 \le M_{U^c L} \le 5000$~GeV, $100 \le M_{E^c} \le 5000$~GeV, $100 \le M_{Q D^c N^c} \le 5000$~GeV, $-5000 \le A_{\tau} \le 5000$~GeV, $-5000 \le A_t \le 5000$~GeV, $-5000 \le A_b \le 5000$~GeV, $2 \le {\rm tan}\beta \le 60$, and $855 \le M_V \le 20,000$~GeV. A reasonable tolerance of the top quark mass is permitted around the world average~\cite{CDF:2013jga}, employing upper and lower limits of $171.8 \le M_t \le 174.8$~GeV. The WMAP 9-year~\cite{Hinshaw:2012aka} and 2015 Planck~\cite{Ade:2015xua} relic density measurements are implemented, such that we constrain the model to satisfy both sets of data and allow the inclusion of multi-component dark matter beyond the neutralino, imposing limits of $\Omega h^2 \le 0.1221$. We will further identify the subset of points that are consistent with the recently released 2018 Planck observation~\cite{Aghanim:2018eyx} of $\Omega h^2 = 0.120 \pm 0.001$. Regarding LHC gluino searches, a firm lower limit is imposed on the gluino mass of $M({\widetilde{g}}) \ge 1.6$~TeV, allowing for a reasonable lower boundary under all the LHC gluino searches.

The light Higgs boson mass theoretical calculation is allowed to float around the experimental central value of $m_h = 125.09$~GeV, though we do enforce a larger range of $123 \le m_h \le 128$~GeV to account for a 2$\sigma$ experimental uncertainty over and above a theoretical uncertainty of 1.5 GeV in our computations. The actual numerical value of the flippon Yukawa coupling remains an unknown, hence we permit the Yukawa coupling to range from a minimum value to its maximum in our light Higgs boson mass computations. A minimal flippon Yukawa coupling does not allow any vector-like flippon contributions, as the light Higgs boson mass in this case is then comprised of only the 1-loop and 2-loop SUSY contributions, chiefly from the coupling to the light stop. A maximum flippon Yukawa coupling implies the $(XD,~XD^c)$ flippon Yukawa coupling is fixed at $Y_{XD} = 0$ and the $(XU,~XU^c)$ flippon Yukawa coupling is set at $Y_{XU} = 1$, with the $(XD,~XD^c)$ flippon trilinear coupling $A$ term set at $A_{XD} = 0$ and the $(XU,~XU^c)$ $A$ term fixed at $A_{XU} = A_U = A_0$~\cite{Huo:2011zt,Li:2011ab}. When necessary, we shall choose the maximum flippon Yukawa coupling in order to lift the light Higgs boson mass up to its observed value. However, some points in the model space can reach the Higgs boson observed value with no flippon contribution, thus we choose the minimal coupling in these instances. Regardless of our choice for the numerical value of the flippon Yukawa coupling, the calculation must return a value within $123 \le m_h \le 128$~GeV. For our chosen benchmarks points that are meant to be sufficiently representative of the model space, we will clearly annotate as to whether the given light Higgs boson mass consists of the minimum or maximum flippon Yukawa coupling.

The model space is additionally constrained via rare decay, direct dark matter detection, and proton decay experimental results. The rare decay experimental constraints include the branching ratio of the rare b-quark decay of $Br(b \to s \gamma) = (3.43 \pm 0.21^{stat}~ ±\pm 0.24^{th} \pm 0.07^{sys}) \times 10^{-4}$~\cite{HFAG}, the branching ratio of the rare B-meson decay to a dimuon of $Br(B_s^0 \to \mu^+ \mu^-) = (2.9 \pm 0.7 \pm 0.29^{th}) \times 10^{-9}$~\cite{CMS:2014xfa}, and the 3$\sigma$ intervals around the Standard Model result and experimental measurement of the SUSY contribution to the anomalous magnetic moment of the muon of $-17.7 \times10^{-10} \le \Delta a_{\mu} \le 43.8 \times 10^{-10}$~\cite{Aoyama:2012wk}. Direct dark matter detection constraints encompass limits on spin-independent cross-sections for neutralino-nucleus interactions established by the Large Underground Xenon (LUX) experiment~\cite{Akerib:2016vxi}, PandaX-II Experiment\cite{Tan:2016zwf}, and XENON100 Collaboration~\cite{Aprile:2018dbl}, and limits on the proton spin-dependent cross-sections by the COUPP Collaboration~\cite{Behnke:2012ys} and XENON100 Collaboration~\cite{Aprile:2013doa}. Finally, we assess our $SU(5) \times U(1)_X$ grand unification against the current limits of about $1.7 \times 10^{34}$~yrs on the proton decay rate $p \to e^+ \pi^0$~\cite{Takhistov:2016eqm}.

A total of 110 million points were randomly sampled in scans implementing the $SU(5) \times U(1)_X$ $M_{\cal F}$ scale boundary conditions, with only about 37,000 of those points surviving the applied constraints. The SUSY mass spectra, relic density, rare decay processes, and direct dark matter detection cross-sections are calculated with {\tt MicrOMEGAs~2.1}~\cite{Belanger:2008sj} utilizing a proprietary mpi modification of the {\tt SuSpect~2.34}~\cite{Djouadi:2002ze} codebase to run flippon and General No-Scale ${\cal F}$-$SU(5)$ enhanced RGEs, employing non-universal soft supersymmetry breaking parameters at the $M_{\cal F}$ scale. Supersymmetric particle decays are computed with {\tt SUSY-HIT~1.5a}~\cite{Djouadi:2006bz}. The Particle Data Group~\cite{Olive:2016xmw} world average for the strong coupling constant is $\alpha_S (M_Z) = 0.1181 \pm 0.0011$ at 1$\sigma$, and we assume a value in this work of $\alpha_S = 0.1184$. 

\begin{figure*}[htp]
       \centering
        \includegraphics[width=1.0\textwidth]{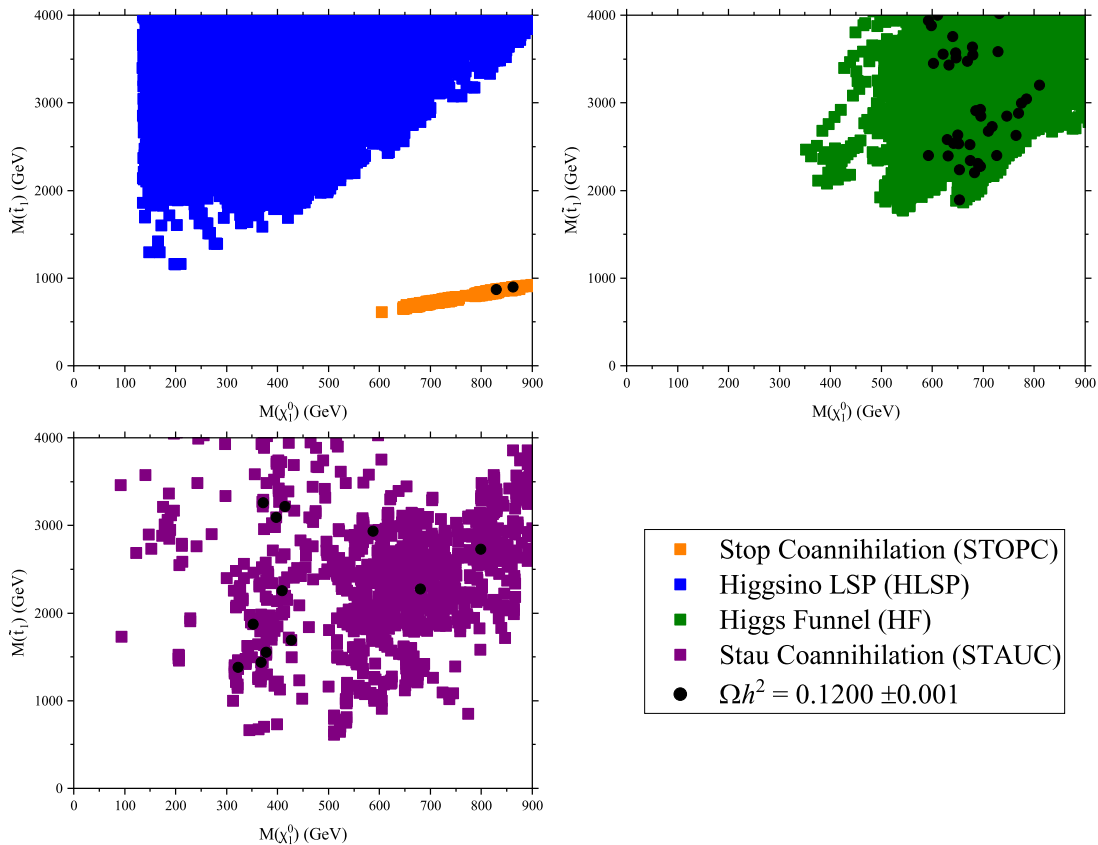}
        \caption{Light stop mass $M(\widetilde{t}_1)$ as a function of the lightest neutralino mass $M(\widetilde{\chi}_1^0)$ for the four regions of the model space we study in this work. The stop coannihilation strip can be observed in the upper left plot space. Notice that there is no intersection between the Light Stop Coannihilation and pure Higgsino regions, though a more detailed analysis relaxing the higgsino and light stop coannihilation requirements we applied here is in progress~\cite{BLMN-P} to ascertain any potential union of these two significant regions. All points plot satisfy the light Higgs boson mass and relic density observations we outlined in this work, in addition to rare decay and current LHC SUSY search constraints. The round black dots represent those points that can also satisfy the recent 2018 Planck Collaboration satellite relic density measurements of $\Omega h^2 = 0.120 \pm 0.001$.}
        \label{fig:stop}
\end{figure*}

\begin{figure*}[htp]
       \centering
        \includegraphics[width=1.0\textwidth]{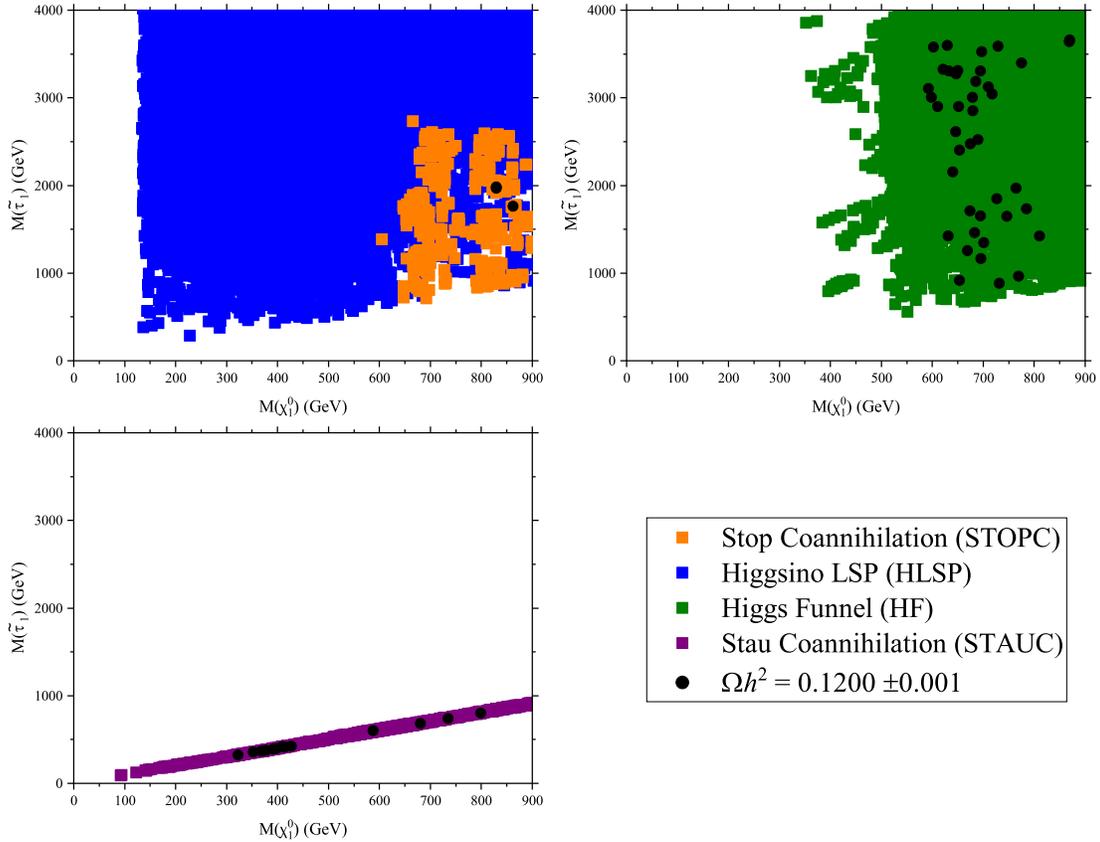}
        \caption{Light stau mass $M(\widetilde{\tau}_1)$ as a function of the lightest neutralino mass $M(\widetilde{\chi}_1^0)$ for the four regions of the model space we study in this work. The stau coannihilation strip can be observed in the lower left plot space. All points plot satisfy the light Higgs boson mass and relic density observations we outlined in this work, in addition to rare decay and current LHC SUSY search constraints. The round black dots represent those points that can also satisfy the recent 2018 Planck Collaboration satellite relic density measurements of $\Omega h^2 = 0.120 \pm 0.001$.}
        \label{fig:stau}
\end{figure*}

\begin{figure*}[htp]
       \centering
        \includegraphics[width=1.0\textwidth]{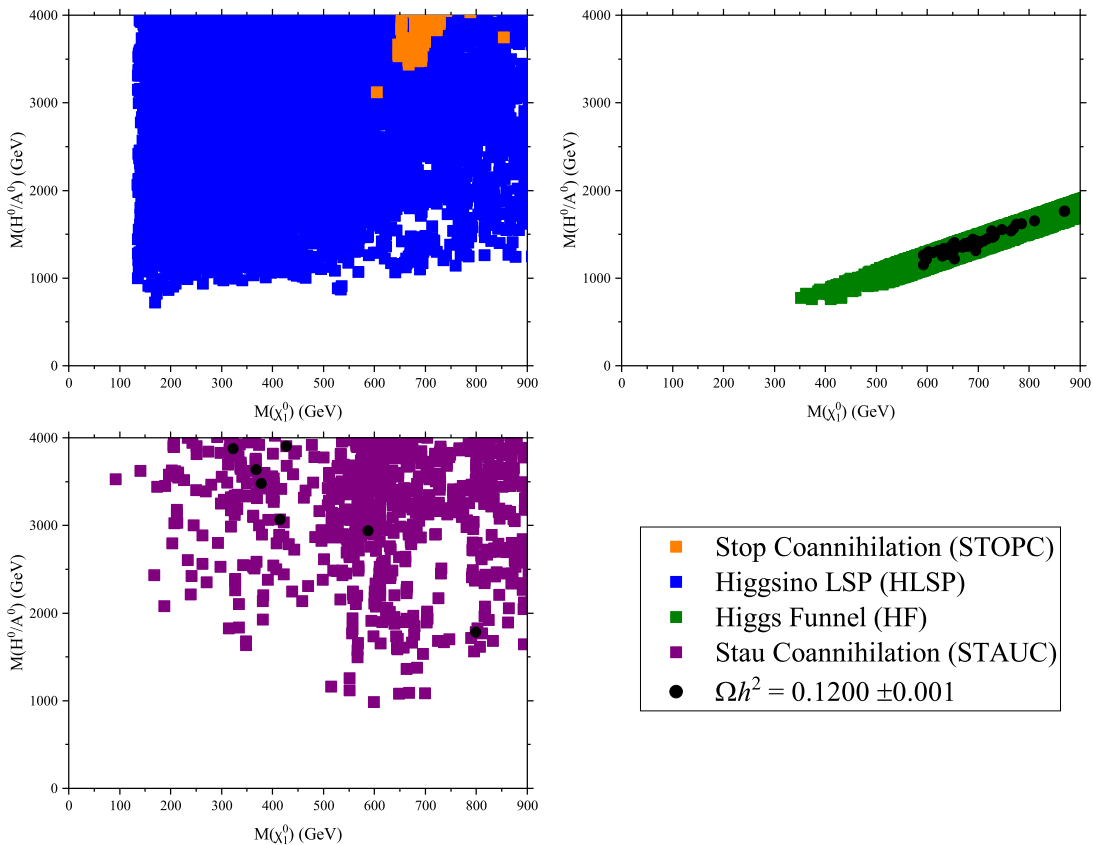}
        \caption{Heavy neutral and pseudoscalar SUSY Higgs mass $M(H^0/A^0)$ as a function of the lightest neutralino mass $M(\widetilde{\chi}_1^0)$ for the four regions of the model space we study in this work. The Higgs Funnel region can be observed in the upper right plot space. All points plot satisfy the light Higgs boson mass and relic density observations we outlined in this work, in addition to rare decay and current LHC SUSY search constraints. The round black dots represent those points that can also satisfy the recent 2018 Planck Collaboration satellite relic density measurements of $\Omega h^2 = 0.120 \pm 0.001$.}
        \label{fig:h0}
\end{figure*}

\begin{figure*}[htp]
       \centering
        \includegraphics[width=1.0\textwidth]{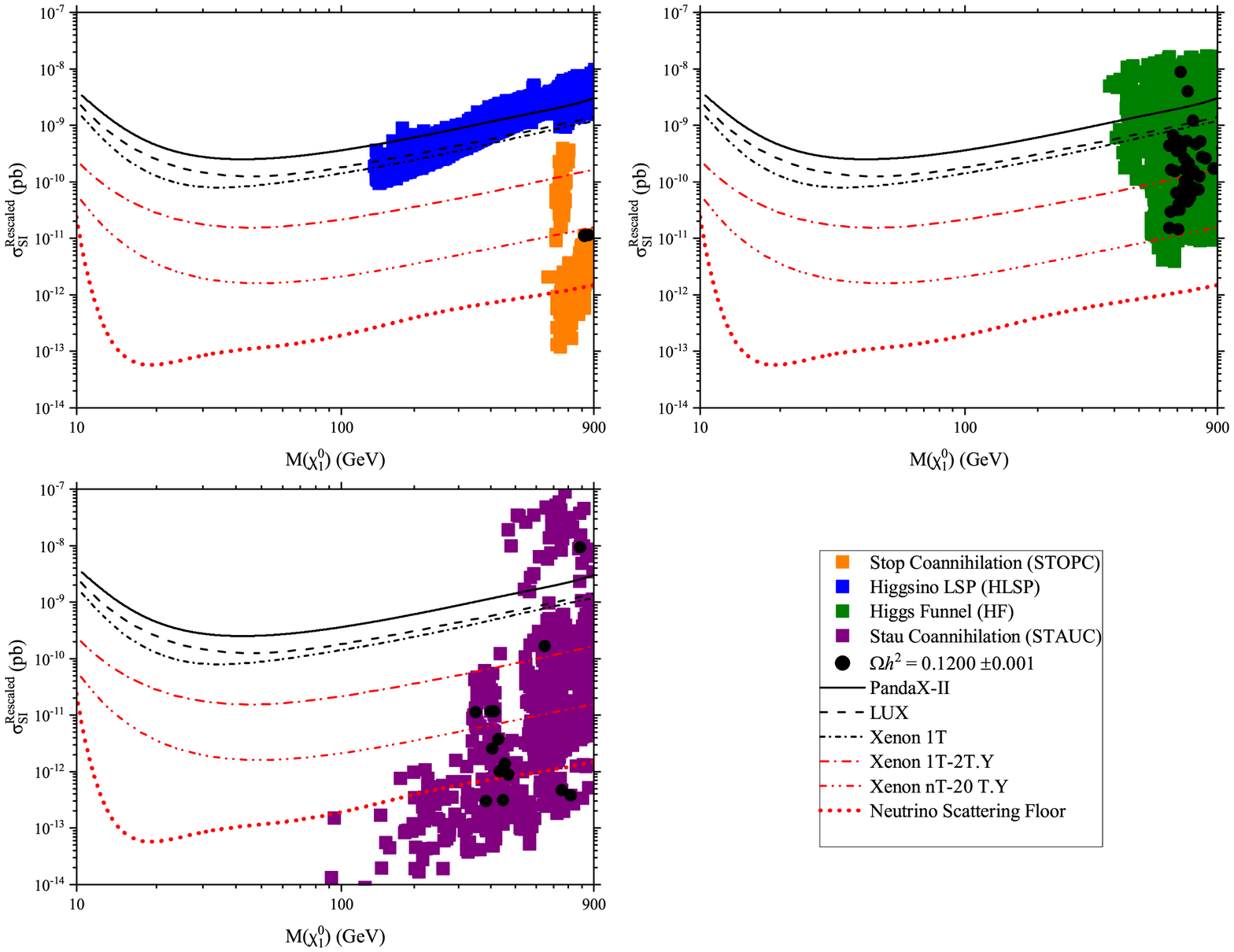}
        \caption{Illustration of the PandaX-II, LUX, and XENON WIMP-nucleon spin-independent cross-section constraints applied to the model space studied in this work. The $\sigma_{SI}$ cross-sections have been rescaled in accordance with Eq.~(\ref{eq:omega}). All points plot satisfy the light Higgs boson mass and relic density observations we outlined in this work, in addition to rare decay and current LHC SUSY search constraints. The round black dots represent those points that can also satisfy the recent 2018 Planck Collaboration satellite relic density measurements of $\Omega h^2 = 0.120 \pm 0.001$.}
        \label{fig:sigma}
\end{figure*}

A total of 12 viable benchmark points are chosen to be reasonably representative of the model space for a given set of parameters ($M_5$, $M_{1X}$, $M_{U^c L}$, $M_{E^c}$, $M_{Q D^c N^c}$, $M_H$, $A_{\tau}$, $A_t$, $A_b$, $M_V$, ${\rm tan}\beta$, $M_t$). The data results are distributed across TABLES~\ref{tab:spectra1} - \ref{tab:spectra3}. The benchmarks models of TABLES~\ref{tab:spectra1} - \ref{tab:spectra3} originate via categorization of the viable model space into four dissimilar regions identified by LSP and NLSP characteristics, to be discussed shortly. The numerical relic density figures annotated in TABLE~\ref{tab:spectra3} consist only of the SUSY lightest neutralino $\widetilde{\chi}_1^0$ abundance, therefore regions with numerical values less than the WMAP9 1$\sigma$ measurement lower bound of about $\Omega h^2 \simeq 0.1093$ must welcome alternative additions to the total observed relic density by WMAP9 and Planck. Given the potential for multi-component dark matter in regions of low neutralino density, the spin-dependent and spin-independent cross-section calculations provided in TABLE~\ref{tab:spectra3} are rescaled per the following expression:
\bea
\sigma^\textrm{re-scaled}_{SI(SD)}=\sigma_{SI(SD)}\frac{\Omega h^2}{0.1138}
\label{eq:omega}
\eea

\section{Phenomenological Results}

The model space is constrained per the experimental constraints detailed in the prior section, with the exception of the LUX, PandaX-II, and XENON cross-sections, which we shall evaluate independently of all the other empirical measurements. Post application of the constraints, the surviving viable parameter space consists of four interesting regions segregated by LSP characteristics and the NLSP. The four regions are identified as (i) light stop coannihilation (STOPC), with $M(\widetilde{t}_1) - M(\widetilde{\chi}_1^0) \le 50$~GeV; (ii) pure higgsino LSP (HLSP), restricted to the conditions $M(\widetilde{\chi}_2^0) - M(\widetilde{\chi}_1^0) = 9-10$~GeV, $M(\widetilde{\chi}_1^{\pm}) - M(\widetilde{\chi}_1^0) = 3-7$~GeV, $M(\widetilde{\chi}_2^0) < 0$, which consistently generates an LSP greater than 98\% higgsino; (iii) Higgs Funnel (HF), with $M(H^0/A^0) \sim 2M(\widetilde{\chi}_1^0)$, where we allow a tolerance of $\pm100$~GeV on $M(H^0/A^0)$; and (iv) light stau coannihilation (STAUC), with $M(\widetilde{\tau}_1^{\pm}) - M(\widetilde{\chi}_1^0) \le 20$~GeV. The light stop coannihilation, Higgs Funnel, and light stau coannihilation regions all generate an LSP that is almost entirely bino. The lower mass boundaries of all these four regions are essentially defined by the LHC constraints, which include the gluino and other ATLAS and CMS search constraints we shall discuss in the forthcoming paragraphs.

The light stop $(\widetilde{t}_1)$ is the NLSP for all of the stop coannihilation points. The very narrow 2018 Planck relic density of $\Omega h^2 = 0.120 \pm0.001$ can be generated with $M(\widetilde{t}_1) - M(\widetilde{\chi}_1^0) \simeq 35-40$~GeV. The STOPC region is illustrated in FIGs.~\ref{fig:stop} - \ref{fig:sigma}, with FIG.~\ref{fig:stop} displaying the light stop mass as a function of the LSP $\widetilde{\chi}_1^0$, clearly indicating the light stop coannihilation strip, with the lower end severed by the most recent LHC constraints. Given that the small mass delta we require is less than $M_t$, the light stop is forced into the decay channel $\widetilde{t}_1 \to c + \widetilde{\chi}_1^0$ and the gluino will always produce a light stop $(\widetilde{g} \to \widetilde{t}_1 t)$, as highlighted in TABLE~\ref{tab:br}. The near degenerate LSP and light stop in this STOPC channel can evade the current ATLAS~\cite{Aaboud:2017phn,Aaboud:2018zjf} and CMS~\cite{Sirunyan:2017kiw,Sirunyan:2017kqq,Sirunyan:2017wif,Sirunyan:2018vjp} LHC light stop searches. Applying the strongest constraints from both ATLAS and CMS on $M(\widetilde{t}_1) - M(\widetilde{\chi}_1^0)$ and $M(\widetilde{t}_1)$ requires us to enforce $M(\widetilde{t}_1) > 550$~GeV for our given mass delta of $M(\widetilde{t}_1) - M(\widetilde{\chi}_1^0) \le 50$~GeV, defining the lower end of the light stop coannihilation strip in FIG.~\ref{fig:stop}. This constraint is included in all of FIGs.~\ref{fig:stop} - \ref{fig:sigma}.

The higgsino spectra are identified by the Higgs bilinear mass term $\mu$ driven via RGE running below the gaugino mass terms $M_1$ and $M_2$ at low-energy near electroweak symmetry breaking (EWSB), thus driving the $\widetilde{\chi}_2^0$ mass negative. This is despite the fact that the $\mu$ term can be larger or smaller than the gaugino mass terms $M_5$ and $M_{1X}$ at the $M_{\cal F}$ scale. The tight constraints applied within the higgsino region ensure an almost 100\% higgsino LSP, though it apparently does prevent any combination of a pure higgsino LSP $and$ light stop NLSP. This is evident in FIG.~\ref{fig:stop} by observing the gap between the HLSP and STOPC points, as no spectra with $M(\widetilde{t}_1) - M(\widetilde{\chi}_1^0) \le 50$~GeV can also generate a pure higgsino LSP. This notwithstanding, preliminary studies do show that a STOPC point with $M(\widetilde{t}_1) - M(\widetilde{\chi}_1^0) \le 50$~GeV can possess an LSP that is dominantly higgsino, for example, more than 60\% and potentially as high as 80\% or greater. While we do not integrate this more detailed HLSP+STOPC analysis into this work, it is currently in progress~\cite{BLMN-P}. Such a combination of a dominantly higgsino LSP and light stop NLSP is rather difficult to produce naturally~\cite{Ellis:2018jyl}, though it has been uncovered in certain realistic intersecting D6-brane models~\cite{Ahmed:2017ttx}. As a result of our tight constraints on the higgsino in order to generate a pure higgsino LSP, most of the higgsino points in FIGs.~\ref{fig:stop} - \ref{fig:sigma} have $M(\widetilde{t}_1) > M(\widetilde{g})$, thus we focus on these primary higgsino spectra here in this work. Unfortunately, as depicted in TABLE~\ref{tab:br}, there is no dominant decay channel to consistent final states for either the light stop or gluino in the HLSP region. As can be inferred from TABLE~\ref{tab:br} and our constraint of $M(\widetilde{\chi}_1^{\pm}) - M(\widetilde{\chi}_1^0) = 3-7$~GeV, the chargino participates in the off-shell Standard Model boson mediated decays $\widetilde{\chi}_1^{\pm} \to u\bar{d}/c\bar{s} + \widetilde{\chi}_1^0$ (36\%/36\%) or $\widetilde{\chi}_1^{\pm} \to l^{\pm} + \nu_l + \widetilde{\chi}_1^0$ (28\%), and these pure higgsino points have an LHC established lower bound of about $M(\widetilde{\chi}_1^{\pm}) \sim 140$~GeV~\cite{Aaboud:2017leg}, thus we enforce $M(\widetilde{\chi}_1^{\pm}) > 140$~GeV in FIGs.~\ref{fig:stop} - \ref{fig:sigma}.

\begin{table*}
\caption{Dominant gluino ($\widetilde{g}$) and light stop ($\widetilde{t}_1$) decay modes for each of the regions of the model space we study in this work, along with the associated branching ratios. Here, $q = (u,d,c,s)$ and $\widetilde{q} = (\widetilde{u},\widetilde{d},\widetilde{c},\widetilde{s})$.}
\label{tab:br}
\begin{tabular}{c c c}
\hline
${\rm Model}$ & ${\rm Dominant~Decay~Mode}$ & ${\rm Branching~Ratio}$ \\ \hline
$$ & $$ & $$ \\
$$ & $\widetilde{g} \to \widetilde{t}_1 t$ & $100\%$ \\
${\rm Stop~Coannihilation}$ & $$ & $$ \\
$$ & $\widetilde{t}_1 \to c + \widetilde{\chi}_1^0$ & $80-98\%$ \\ 
$$ & $$ & $$ \\ \hline
$$ & $$ & $$ \\
$$ & $\widetilde{g} \to \widetilde{\chi}_1^{\pm} + t + b \to q \bar{q} + t + b + \widetilde{\chi}_1^0$ & $\sim 28\%$ \\
$$ & $\widetilde{g} \to t \bar{t} + \widetilde{\chi}_1^0$ & $\sim 14\%$ \\
${\rm Pure~Higgsino}$ & $$ & $$ \\
$$ & $\widetilde{t}_1 \to \widetilde{g} + t $ & $\sim 32\%$ \\
$$ & $\widetilde{t}_1 \to \widetilde{\chi}_1^{\pm} + b \to q \bar{q}  + b +   \widetilde{\chi}_1^0$ & $\sim 16\%$ \\ 
$$ & $$ & $$ \\ \hline
$$ & $$ & $$ \\
$$ & $\widetilde{g} \to \widetilde{\chi}_1^{\pm} + t + b \to q \bar{q} + t + b +  \widetilde{\chi}_1^0~~\left[ M(\widetilde{\chi}_1^{\pm}) - M(\widetilde{\chi}_1^0) < M(W^{\pm}) \right]$ & $\sim 21\%$ \\
$$ & $\widetilde{g} \to \widetilde{\chi}_1^{\pm} + t+ b \to W^{\pm} + t + b +  \widetilde{\chi}_1^0~~\left[ M(\widetilde{\chi}_1^{\pm}) - M(\widetilde{\chi}_1^0) > M(W^{\pm}) \right]$ & $\sim 26\%$ \\
$$ & $$ & $$ \\ 
$$ & $\underline{{\rm Larger}~M(\widetilde{t}_1) - M(\widetilde{g})}$ & $$ \\
$$ & $\widetilde{t}_1 \to \widetilde{g} + t $ & $\sim 48\%$ \\
$$ & $\widetilde{t}_1 \to \widetilde{\chi}_1^{\pm} + b \to q \bar{q} + b +  \widetilde{\chi}_1^0$ & $\sim 15\%$ \\
${\rm Higgs~Funnel}$ & $$ & $$ \\
$$ & $$ & $$ \\
$$ & $\underline{{\rm Smaller}~M(\widetilde{t}_1) - M(\widetilde{g})}$ & $$ \\
$$ & $\widetilde{t}_1 \to \widetilde{g} + t $ & $\sim 18\%$ \\
$$ & $\widetilde{t}_1 \to \widetilde{\chi}_1^{\pm} + b \to W^{\pm} + b +  \widetilde{\chi}_1^0$ & $\sim 39\%$ \\
$$ & $$ & $$ \\ \hline
$$ & $$ & $$ \\
$$ & $\underline{M(\widetilde{g}) - M(\widetilde{t}_1) > M_t}$ & $$ \\
$$ & $\widetilde{g} \to \widetilde{t}_1 t \to t \bar{t} +  \widetilde{\chi}_1^0$ & $\sim 100\%$ \\
$$ & $\widetilde{t}_1 \to t +  \widetilde{\chi}_1^0$ & $\sim 100\%$ \\
${\rm Stau~Coannihilation}$ & $$ & $$ \\ 
$$ & $\underline{M(\widetilde{g}) - M(\widetilde{t}_1) < M_t}$ & $$ \\
$$ & $\widetilde{g} \to t \bar{t} +  \widetilde{\chi}_1^0$ & $\sim 6\%$ \\
$$ & $\widetilde{t}_1 \to \widetilde{\chi}_1^{\pm} + b \to \widetilde{\tau}_1^{\pm} + \nu_{\tau} + b \to \tau^{\pm}  + \nu_{\tau} + b + \widetilde{\chi}_1^0$ & $\sim 18\%$ \\
$$ & $$ & $$ \\ \hline
\end{tabular}
\end{table*}

The chargino $\widetilde{\chi}_1^{\pm}$ is the NLSP for 99.6\% of the Higgs Funnel points, with the remaining 0.4\% possessing a stau NLSP. Nonetheless, only 0.07\% of the Higgs Funnel also resides in the $M(\widetilde{\tau}_1^{\pm}) - M(\widetilde{\chi}_1^0) \le 20$~GeV stau coannihilation strip. We further observe that 11.5\% of the Higgs Funnel has a dominant higgsino LSP, though only 0.5\% are pure higgsino. The upper right plot space of FIG.~\ref{fig:h0} clearly shows the HF region with $M(H^0/A^0)$ plot as a function of the LSP $\widetilde{\chi}_1^0$. The minimum heavy Higgs boson mass in the HF region is $M(H^0/A^0) \sim 800$~GeV, and these $H^0/A^0$ near this lower boundary all have $30 \le {\rm tan}\beta \le 40$, and given a branching ratio of more than 85\% for $H^0/A^0 \to b \bar{b}$, the lightest $M(H^0/A^0)$ in our model space persist comfortably beyond current LHC constraints for BSM Higgs searches~\cite{Sirunyan:2018taj}.

In our light stau coannihilation strip, only 27.6\% of the points have stau NLSP, with the remaining 72.4\% having a chargino $\widetilde{\chi}_1^{\pm}$ NLSP. As one would expect though, four nearly degenerate sparticles ($\widetilde{\chi}_1^{0}$, $\widetilde{\chi}_1^{\pm}$, $\widetilde{\chi}_2^{0}$, $\widetilde{\tau}_1^{\pm}$) greatly suppresses the relic density, hence only the stau $\widetilde{\tau}_1^{\pm}$ NLSP points with larger $\widetilde{\chi}_1^{\pm}$ and $\widetilde{\chi}_2^{0}$ in the light stau coannihilation strip can generate the observed WMAP9 and Planck relic density. The light stau coannihilation strip is evident in the lower left panel of FIG.~\ref{fig:stau}. We inspect the most recent ATLAS slepton constraints~\cite{Aaboud:2017leg} on compressed spectra, such as we have in the STAUC region, therefore we attempt to approximately emulate the ATLAS slepton exclusions by applying the following constraints on our model space, where both conditions for each must be TRUE for the point to be excluded: (1) $M(\widetilde{\tau}_1^{\pm}) < 100$~GeV and $M(\widetilde{\tau}_1^{\pm}) - M(\widetilde{\chi}_1^0) > 1$~GeV; (2) $100  \le M(\widetilde{\tau}_1^{\pm}) < 150$~GeV and $2  \le M(\widetilde{\tau}_1^{\pm}) - M(\widetilde{\chi}_1^0) \le 14$~GeV; and (3) $150  \le M(\widetilde{\tau}_1^{\pm}) < 190$~GeV and $3  \le M(\widetilde{\tau}_1^{\pm}) - M(\widetilde{\chi}_1^0) \le 8$~GeV. These restrictions are included in FIGs.~\ref{fig:stop} - \ref{fig:sigma}. Notice in constraint (1) just above that we allow for an off-shell tau lepton $\tau^{\pm}$ less than 1~GeV, an action that does capture a not insignificant number of viable points, in view of the fact that 10.5\% of the STAUC region has $M(\widetilde{\tau}_1^{\pm}) - M(\widetilde{\chi}_1^0) < 1$~GeV. A quick review of TABLE~\ref{tab:br} indicates that the stau coannihilation spectra with light stau NLSP and $M(\widetilde{g}) - M(\widetilde{t}_1) > M_t$ produce exclusively a 4-top signature ($\widetilde{g} \to \widetilde{t}_1 t \to t \bar{t} +  \widetilde{\chi}_1^0$), giving rise to large multijet events, as is typical in \fsu5~\cite{Maxin:2011hy}. The final states are not as clear when $M(\widetilde{g}) - M(\widetilde{t}_1) < M_t$ or $M(\widetilde{g}) < M(\widetilde{t}_1)$ given that there is no dominant decay channel in either of these two cases. The model space is generally split between these two situations, where about half the STAUC region has $M(\widetilde{g}) - M(\widetilde{t}_1) > M_t$ and the other half of the STAUC region has $M(\widetilde{g}) - M(\widetilde{t}_1) < M_t$ or $M(\widetilde{g}) < M(\widetilde{t}_1)$.

\section{Conclusion}

With our motivation partially inspired by D-brane model building, we presented an analysis of non-zero general SUSY breaking soft terms in ${\cal F}$-$SU(5)$. The methodology involved massive parallel computing given the large number of unknown parameters. While the resulting viable parameter space was extraordinarily constrained given the large number of computations, several interesting regions were uncovered that can simultaneously generate the WMAP and Planck observed relic density measurements and correct light Higgs boson mass, over and above satisfying many LHC search constraints. Four regions that accomplish these rely upon (i) light stop coannihilation, (ii) pure Higgsino dark matter, (iii) Higgs funnel, and (iv) light stau coannihilation, though the pure Higgsino LSP could not reach the observed relic density due to its large annihilation cross-section. Concluding the analysis was an effort to identify the decay modes to final states that could represent observable signatures of these four scenarios, discovering that the light stop coannihilation and light stau coannihilation decay channels possess very dominant branching fractions. The light stop coannihilation produces the typical top+charm quark final state via the gluino decay and the light stau coannihilation mostly leads to the characteristic \fsu5 large multijet event, whereas the pure Higgsino LSP and Higgs funnel provide no dominant decay channel to final states. Though possibly furnishing the highest intrigue, our exploration tentatively revealed that the \fsu5 model may indeed harbor a rather diminutive region exhibiting spectra with a mixed scenario of a dominant Higgsino LSP $and$ light stop coannihilation, a rare yet rather natural SUSY spectrum, but that is the focus of our next endeavor.
 

\section{Acknowledgments}

Portions of this research were conducted with high performance computational resources provided 
by the Louisiana Optical Network Infrastructure (http://www.loni.org). This research was supported 
in part by the Projects 11475238, 11647601, and 11875062 supported 
by the National Natural Science Foundation of China (TL), and by the DOE grant DE-FG02-13ER42020 (DVN). 


\bibliography{bibliography}

\end{document}